# Dorsal and ventral target strength measurements on gilthead sea bream (*Sparus aurata*) in sea cages

Ester Soliveres, Alejandro Cebrecos, Víctor Espinosa

Instituto de Investigación para la Gestión Integrada de Zonas Costeras (IGIC). Universidad Politécnica de Valencia. c/ Paranimf, 1, 46730, Grao de Gandia, Valencia, España. (essogon, alcebrui, vespinos)@upv.es

Corresponding author: Tel.: +34 963 877 000 Ext. 43680/1; Fax: +34 962 849 337; email: essogon@upv.es; address: I. Inv. Gestión Integrada de Zonas Costeras, Universidad Politécnica de Valencia, c/ Paranimf nº 1, 46730, Grao de Gandia, Valencia, Spain

## Abstract

The aim of this study is to establish a relationship between target strength (*TS*) and total body length of the gilthead sea bream (*Sparus aurata*), in order to monitor its growth in sea cages. Five classes of commercial size gilthead sea bream are characterized, comprising lengths from 20 to 25 cm, corresponding to weights between 160 and 270 g. A few specimens were introduced into a sea cage of 3 m in diameter and a height of 2.7 m. We measure *TS* directly using a Simrad EK60 echosounder with a 7º split-beam transducer working at 200 kHz. The transducer was located in the center of the cage during measurements, at the bottom facing upwards for ventral recordings and on the surface facing downwards to perform dorsal recordings.

Two analyses based on single echo detection were performed: the first one obtains compensated transducer directivity *TS* values from intensity and angular echosounder data; while the second one omit phase information, affording uncompensated *TS* values (*TSu*). Two algorithms have been applied to analyze single-beam-like data, that differ in the order in which threshold and echo length criteria are applied to detect single echoes. *TS* distributions obtained from the split-beam analysis are unimodal for both ventral and dorsal measurements, like *TSu* distributions obtained when the threshold criterion is applied first. A lineal relationship was found between mean *TS* for the ventral aspect and logarithm of total body length of fish, showing good correlations for both *TS* and *TSu*, even having a few detections. Note that the relationship between *TSu* and total body length allows monitoring the growth of gilthead sea bream using low cost equipment like single beam echosounders.

## 1 Introduction

In recent years, marine aquaculture has increased considerably due to the lack of fishery resources and the growing demand for marine products. Aquaculture currently accounts for 50% of global aquatic production for human consumption. In Spain, the third country in production of farmed fish of the European Union, this increase in aquaculture production is due mainly to farming of gilthead sea bream (*Sparus aurata*) and sea bass (*Dicentrarchus labrax*), representing 95% of total farmed fish production. The gilthead sea bream is one of the most interesting species for farming in the Mediterranean basin, the third most produced fish species in the EU and the first in Spain. It is grown in a floating cage system and its production is limited to areas in temperate waters (APROMAR, 2009; FAO, 2010).

Growth rate and biomass estimation of fish are essential in aquaculture, in order to prepare the production plan of fish farms, as well as management tasks such as classification and distribution of fish, discharge of new lots, calculation of daily feeding rates, etc. Despite the fact that existing technology has sufficient capacity to satisfy production needs, it is necessary to optimize these processes, not only to improve economic efficiency, but also to minimize the environmental impact of fish farms. These processes include feeding strategy, growth rate and population monitoring. Daily feeding is estimated according to present biomass in the sea cage and several factors such as average size of fish, season, water temperature, etc. Therefore, size and fish number estimation in each sea cage are crucial for suitable management of fish farms.

Several surveys related to *in situ* measurements of dorsal aspect of scattering properties of the wild fish in fisheries applications can be found in the literature. McClatchie et al. (1996) analyzed results obtained using *in situ* techniques in multiple studies on the relationship between target strength (*TS*) and total body length for different species of fish, mainly swimbladdered ones. Furthermore, many studies were conducted using *ex situ* techniques with tethered fish or confined in very small cages (Kang and Hwang, 2003, Kang et al., 2004; Kang et al., 2009; Gauthier and Rose, 2001; Mukai and Iida, 1996; Nielsen and Lundgren, 1999; Lilja et al., 2000), in order to determine directivity of the energy scattered by the fish. However, a small number of studies with captive fish have been performed in sea cages for aquaculture applications, mainly with mackerel, herring and capelin (Edwards and Amstrong, 1983; Ona, 2003; Jorgensen and Olsen, 2002; Jorgensen, 2003), and more recently the acoustic scattering of salmon has been studied, measured for both dorsal and ventral aspect (Knudsen et al., 2004). Currently, there are no studies on the backscattering of gilthead sea bream, however other species of the same family have been studied using fish tied or placed in very small cages (Kang and Hwang, 2003; Kang et al., 2004).

Fish behavior (tilt angle, swimming depth, activity level, etc.) greatly influences *TS* values (Ona, 1999; Simmonds and MacLennan, 2005; Medwin, 2005). It is therefore desirable to perform measurements in a similar environment where the biomass is being estimated, since fish behavior is closer to the real case. In aquaculture applications it seems very advisable to measure the backscattering energy of gilthead sea bream inside a sea cage, rather than employing tethered fish or placed in very small cages. The measuring cage must have similar dimensions to a marine fish farm cage, where fish can swim freely. Nevertheless this can be difficult in production conditions since the density of fishes is extremely high (up to 25 kg/$m^3$) and single echo o trace detection can be achievable only in the closest distances from the transducer (up to 3 m). Therefore it can be also important to measure the *TS* at short ranges in a small sea cage. The measurements made in sea cages present some complications, since the fish are very close to the transducer. Note that the near field of a fish can reach several meters, so that *TS* may depend on the distance and, therefore, *TS* measurements in sea cages may not be valid to be applied to biomass estimation in the wild (Ona, 1999). At short distances fishes do not act as point sources, so it will be necessary to estimate the appropriate TVG function to obtain the absolute *TS* values properly (Dawson et al., 2000; Mulligan, 2000). In addition, there is considerable variability in the estimation of the position of fish within the beam at short distances since fish are complex scatterers, so a new error in *TS* estimation will be introduced (Dawson et al., 2000).

Teleost fish, such as the sea bream, normally have a swimbladder. This organ, located below fish bone, is responsible for most of the energy reflected by fish (Simmonds and MacLennan, 2005; Medwin, 2005). Therefore, ventral *TS* measurements in sea cages are highly recommended, because it is more likely to get better correlation between mean *TS* and fish length than that obtained for dorsal aspect. When dorsal measurements are preformed, the shadowing effect of the swimbladder by bony structures (spine) can occur. Whereas this will probably not happen for ventral measurements (Foote, 1985; Medwin and Clay, 1998).

Although different techniques have been developed for the control of marine farms, permanent installation of commercial equipment in sea cages is difficult because of its high price. Single-beam echosounders can be a convenient option for installation in fish farms, as this type of echosounders are the cheapest system among the existing scientific ones. To confirm this, it is necessary to verify the feasibility of control of growth rate, abundance and feeding of fish in sea cages using this type of system.

The aim of this study is to find a relationship between *TS* and total length of gilthead sea bream in sea cages. Note that as we assume the existence of error sources in measurements, since the fish are very close to the transducer, it is not pretended to perform an absolute determination of *TS*, but to find a relationship that allows monitoring the growth of specimens in sea cages using acoustic techniques. We have evaluated the relationship obtained from data provided by a split-beam echosounder and contrasted it with that obtained by omitting the phase information, i.e., treating the data as if they were measured with a single-beam echosounder. Knowledge of these relationships is very important for the management of farms by using acoustic techniques. The ultimate goal of our project is the evaluation of a low-cost single-beam system that is able to withstand long periods of time on a marine farm in production conditions.

# 2 Material and methods

## 2.1 Experimental measurements

Five different size classes of gilthead sea bream were characterized to find the relationship between dorsal and ventral aspect of *TS* for individual specimens and total length of its body. Both biometric and acoustic measurements took place in July 2009.

Firstly, total body length and total mass of different specimens, previously anesthetized, were measured. Fishes were classified into five different size classes between 20 and 25 cm, trying to minimize size differences between fishes of the same class. Size classification covers only part of production sizes, with fishes from 25 to 500 g, and represents an effort of distinguishing small steps in size evolution. Table 1 summarizes mean mass and mean total length of specimens from each class, variation coefficient for total length and number of individuals inside the sea cage.

***Table 1.*** *Biometric data for each size class of gilthead sea bream.*

| Size class | 1 | 2 | 3 | 4 | 5 |
|---|---|---|---|---|---|
| Number of fish | 2-4 | 2-4 | 2-8 | 2 | 2-4 |
| Mass (g) | 158.8 | 178.7 | 194.4 | 235.7 | 268.7 |
| Total length (cm) | 20.3 | 21.1 | 21.7 | 23.1 | 24.2 |
| CV Length (%) | 0.25 | 0.42 | 0.51 | 0.55 | 0.42 |

Secondly, a direct ventral and dorsal *TS* measurement of single fishes was performed. Measurements were carried out at Gandia’s port (Valencia, Spain), using an experimental cage of 3 m in diameter and 2.7 m in height, where the fish can swim freely. A limited number of specimens from each size class were placed in the cage, in order to avoid echoes from different targets which may overlap and falsify echo detections from single targets, and thus overestimate mean *TS* (Ona, 1999). After entering the fish in the sea cages, we waited about one hour before starting the recordings, so fishes got used to the new environment. Fishes were fed until the day before measurements began; to prevent swimbladder volume variation.

A Simrad EK60 echosounder with a 7º split-beam transducer working at 200 kHz, with a maximum near field of about 1 m, was used for direct measurement of *TS*. The transducer is mounted in the center of the bottom of sea cage oriented upwards for ventral *TS* measurement and on the surface facing downwards for dorsal *TS* measurement (Plate 1). Transmitted power was 90 W, pulse length of 64 µs and ping interval of 20 ms. Water temperature was monitored during recordings. Before placing fish in the sea cage and starting measurements, the transducer was calibrated inside the sea cage using a copper calibration sphere of 13.7 mm in diameter with a *TS* of -47 dB at 200 kHz (Simmonds and MacLennan, 2005; Medwin and Clay, 1998).

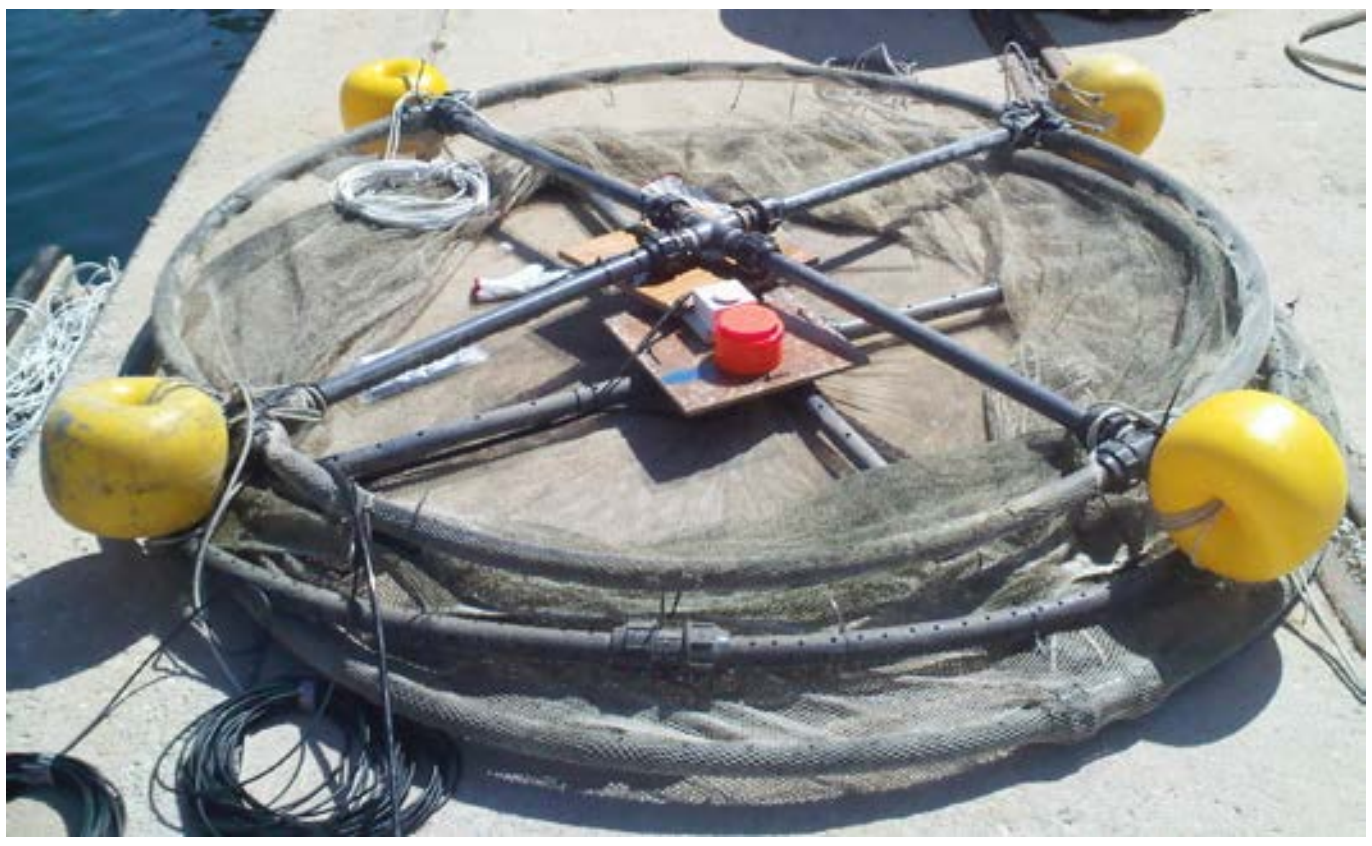

***Plate 1.*** *Mounting detail of the transducer at the bottom of sea cage for ventral TS measurement of the gilthead sea bream.*

Recordings were performed for about an hour for each position of the transducer (ventral and dorsal measurements) and size class. Recorded data were stored in .raw files to later analysis.

## 2.2 Data analysis

Two different data analysis were performed. First one consisted on analyzing all the information provided by the split-beam echosounder, including distance and angle of arrival of received echoes. Detected echo level depends on directivity pattern of the transducer, and angular position information allows obtain compensated *TS*. Second one is a single-beam analysis, which obviates angular information and uncompensated *TS* values (*TSu*), which depend on the arrival direction of the echo, are obtained.

Split-beam data were analyzed using Sonar5-Pro software, version 5.9.9 (Balk and Lindem, 2009). Obtained *TS* distributions are based on analysis of echoes from single targets, since the number of detections is not enough to obtain reliable results from a trace analysis. Firstly, existent noise or unwanted echoes in the echogram are reduced using target-noise separation. Then, a layer that extends from 1 m to 2.35 m from the transducer is analyzed by echo counting. Distances below 1 m are not analyzed to avoid near field effect of the transducer, and over than 2.35 m to avoid echoes coming from surface or bottom of the cage, as well as noise from bubbles due to waves, surface floating objects, etc. It has been necessary, in some cases, to manually remove remaining noise in order to reduce the effect of the inclusion of unwanted echoes in *TS* distributions. A detection is considered to come from single fish if it has a pulse length between 0.8 and 1.8 relative to the duration of transmitted pulse, a maximum gain compensation in one-way of 6 dB, and a maximum phase deviation of 0.5. A threshold of -60 dB was used for ventral recordings and a threshold of -70 dB for dorsal recordings.

Results provided by Sonar Pro were analyzed statistically to obtain mean *TS* for each size class. Data were transformed to linear domain, averaged and the result was transformed to logarithmic domain to obtain mean *TS*. Finally, least-squares linear fit was performed to find the relationship between mean *TS* and logarithm of total body length of fish. Furthermore, *TS* dependence on range has been studied, analyzing mean *TS* values in layers of reduced thickness (10 cm).

Single-beam data were analyzed using Sonar5-Pro and Matlab®. First, data are read from *.raw* file omitting angular information. In this case, analysis by echo counting was also performed to obtain *TS* distributions, but without pre-noise reduction, using the same parameters both in Sonar5-Pro as in Matlab. An algorithm has been developed in Matlab for single echo detection (SED), based on threshold, echo length, and echo spacing criteria (Ona, 1999; Balk and Lindem, 2009). Since noise is not reduced in the echogram, analysis layer is narrower (1 to 2 m) to try to minimize the number of unwanted echoes that are included in analysis. Threshold value is set at -50 dB, echo length admitted is bounded between 0.8 and 1.8 on the transmitted pulse length, and minimum echo spacing is equal to the transmitted pulse length.

Two different versions of single-beam processing application were implemented. The difference between them lie in the order in which SED criteria are applied, particularly threshold (echo amplitude) and echo length (echo duration) criteria. If echo length criterion is applied before threshold criterion, in a similar manner to Sonar5-Pro (Balk and Lindem, 2009), some of single echoes detected may have less length than fixed. However, if threshold criterion is applied first and then echo length criterion, as suggested by Ona (1999), all single echoes detections will have a length equal to the specified. In this case, high thresholds may cause the removal of low-level echoes, which come from targets located outside the axis of the transducer.

As in the previous analysis, results are statistically analyzed to obtain mean *TSu*, which is averaged in linear domain, for each size class. Least-squares linear fit is performed to find the relationship between *TSu* and logarithm of total length of fish, both dorsal and ventral measurements.

# 3 Results and Discussion

## 3.1 *TS* dependence with range

Fig. 1 shows *TS* values of each detection as a function of range from the transducer for ventral measurements of size class 5, which corresponds to largest fish class, and mean *TS* for layers of 10 cm thick. Mean *TS* of layer tend to mean value of all detections when the analyzed layer has a large number of detections, while it varies greatly in layers having few detections. Fig. 1 shows that there is no clear dependence of *TS* with range.

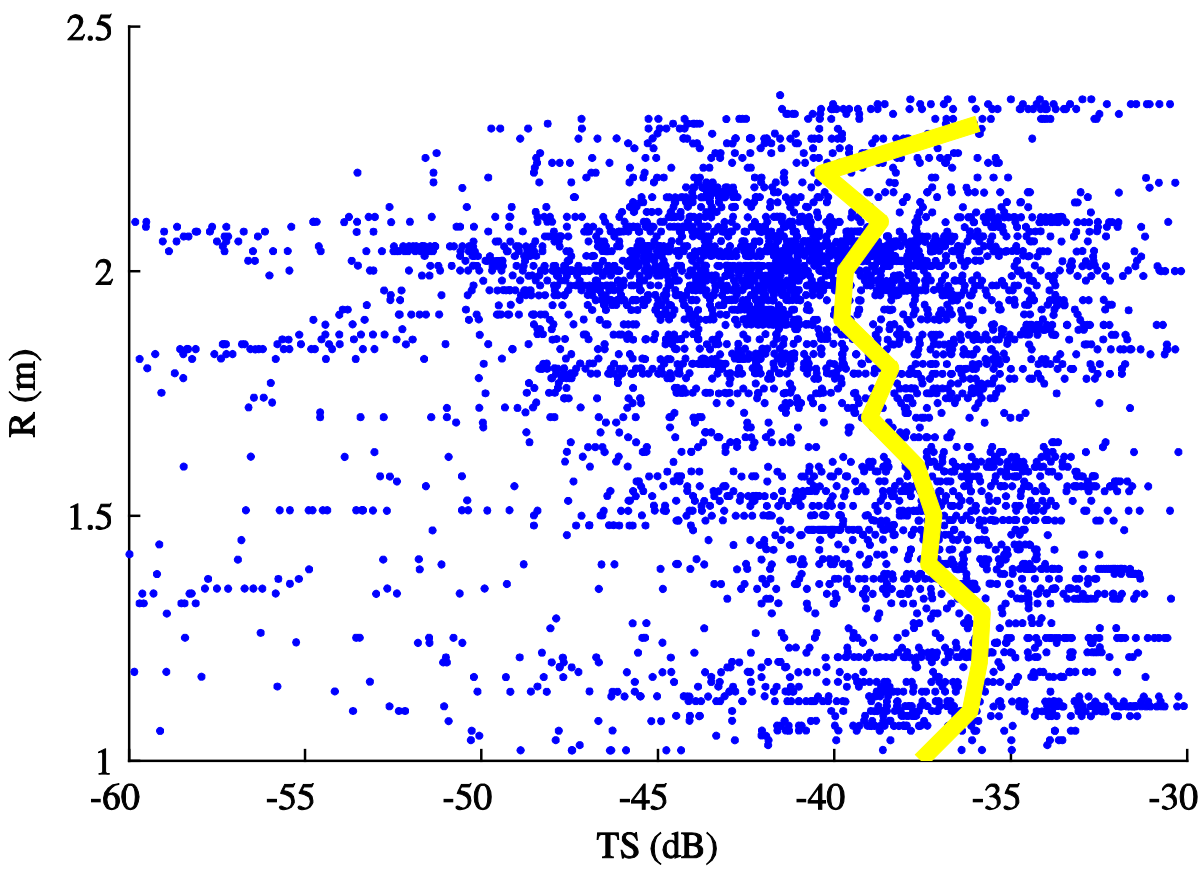

***Fig. 1.*** *TS values as a function of range from the transducer (blue points) and mean TS for layers of 10 cm thick (yellow line) for ventral measurements of the largest size class (class 5).*

Gilthead sea bream is a physoclist fish and, therefore, it has a closed swimbladder able to adjust its volume through glands that allow gas exchange with the outside (Webb, et al., 2008). Some studies about physoclist fish suggest that *TS* don't have dependence with depth at the shallowest range, because they can vary their swimbladder volume as a function of depth, thus maintaining neutral buoyancy (Gauthier and Rose, 2002). Therefore, results are consistent with expected non-dependence of *TS* with depth, typical of physoclist fish.

However, we expected to find some dependence of *TS* with distance due to several reasons: fishes are partially insonificated, and measurements are made in near field of fish (Dawson et al., 2000; Knudsen et al., 2004; Mulligan, 2000). At short distances, fishes are not completely insonificated by the transducer beam, so they do not behave like a point target. For this reason, errors in angle estimation and loss compensation are introduced. Furthermore, near field of a fish can reach several meters, which can cause a large variability in *TS* values recorded and invalidate *TS* measurements in sea cages (Dawson et al., 2000; Knudsen et al., 2004).

Despite the above reasons we have not been observed a clear trend of variation in mean *TS* with range, but a strong dependence on mean *TS* with number of detections in the layer was found. These mean *TS* variations also could be due to near field effect and point source violation. To be able to specify more about these effects in *TS* measurements in sea cages it would be necessary to empirically measure near field of fish, and to estimate angular position error of target and error in loss compensation.

## 3.2 Relationship between *TS* and total body length

### 3.2.1 Split-beam analysis

Fig. 2 shows ventral distribution *TS* at 200 kHz for both the smallest size class (class 1), which has a mean total body length of 20.3 cm, and the second largest size class (class 4), with a mean total body length of 23.1 cm. Both dorsal and ventral measurements show unimodal *TS* distributions. These distributions could be explained by the reduced directivity that examined specimens present, since swimbladder size is only slightly greater than wavelength.

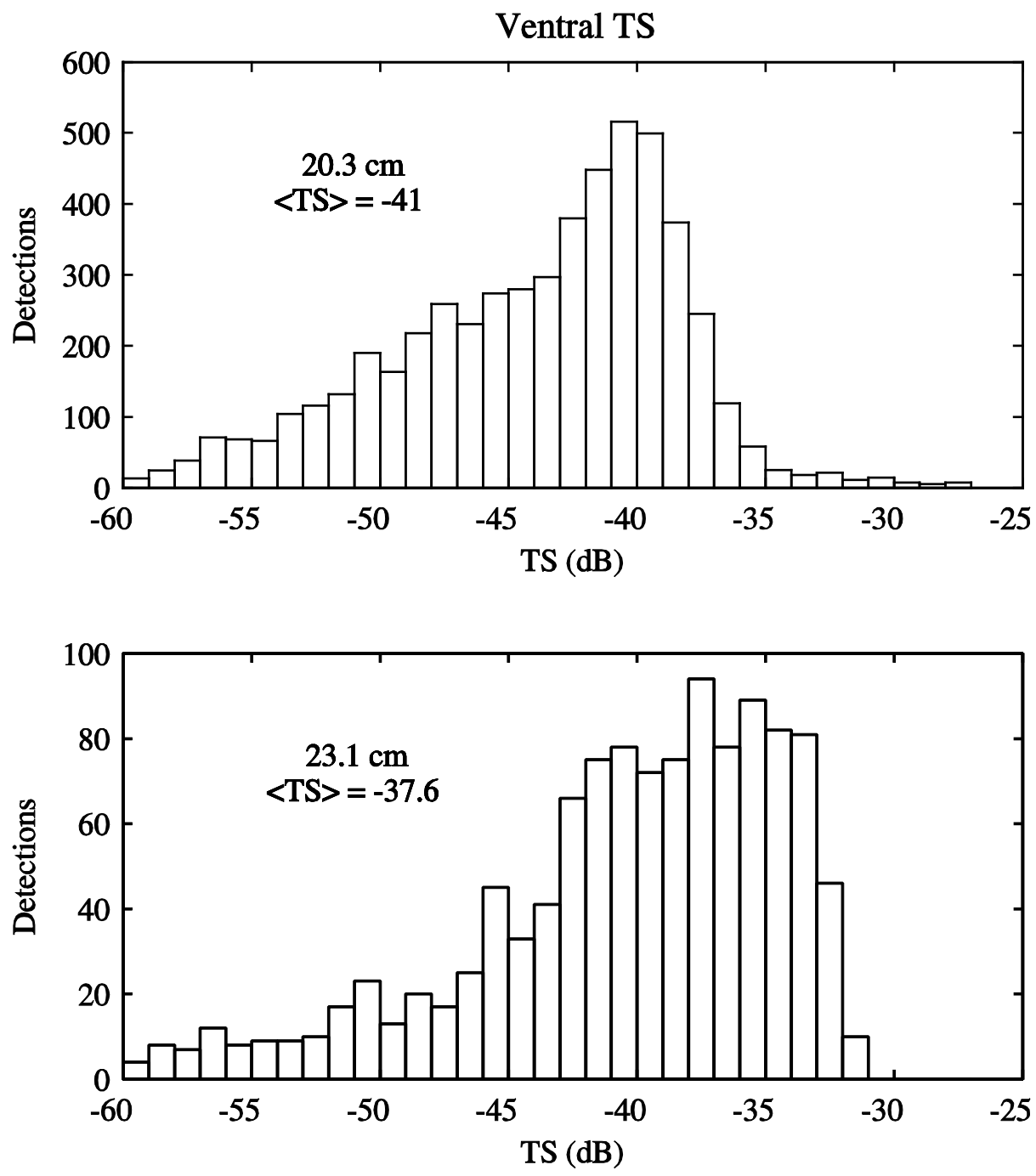

***Fig. 2.*** *Ventral TS distributions for the smaller size class (top), and the second largest class (bottom).*

A clear linear relationship is found between ventral *TS* and logarithm of total body length of gilthead sea bream. Nevertheless, mean *TS* vary greatly between consecutive size classes for dorsal measurements, maintaining a non-monotonic relationship with fish body length (Fig. 2). Mean *TS* for ventral measurements is higher than that obtained for dorsal measurements, like other species of fish (Knudsen et al., 2004), possibly due to shadowing effect of swimbladder by hard structures of fish.

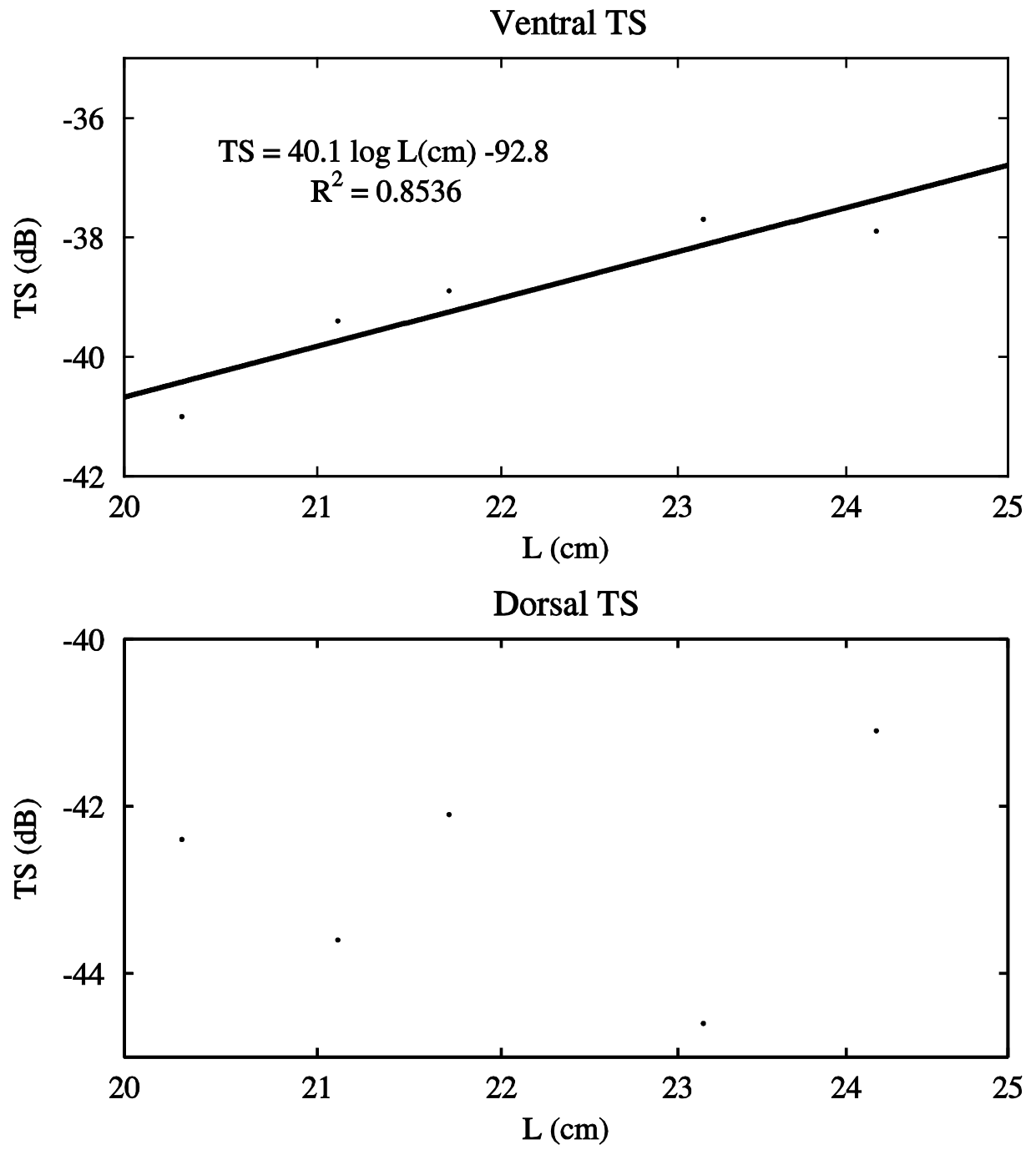

***Fig.3***. *Relationship between mean TS and total body length (in centimeters) of gilthead sea bream, for both ventral (top) and dorsal measurements (bottom).*

Trying to determine the relationship between ventral *TS* and total body length of gilthead sea bream, we noted that the analysis done for class 5 using different parameters and criteria have always shown a lower result than expected. The general trend of the other classes is a linear *TS* increase with logarithm of total body length for all cases. We observed that mean *TS* of class 5 was affected by strong dispersion induced by low *TS* echoes or noise of unknown origin that is not present on the other classes. Mean *TS* of class 5 tends to increase slightly when these echoes are removed from the echogram.

The following linear relationship with logarithm of total body length is established for ventral aspect at a frequency of 200 kHz, from the obtained values of mean *TS*,

$$TS = 40.1 \cdot \log L \text{ (cm)} - 92.8 \qquad (1)$$

$r^2 = 0.8536$

### 3.2.2 Single-beam analysis

Fig. 4 shows the results for ventral measurements of size class 4 from single-beam analysis, where phase information has been obviated. Ventral *TSu* distribution obtained using the processed version which works similarly to Sonar5-Pro, applying echo length criterion firstly, is showed at the top. This distribution is compared with the one obtained using Sonar 5 Pro with the same values of analysis parameters. Ventral *TSu* distribution obtained when data is processed using the other version which applies primarily the threshold criterion, is showed at the bottom, and is also compared with the results obtained using Sonar 5 Pro. In the first case, similar distributions to those obtained with Sonar5-Pro are obtained for all size classes studied, both dorsal and ventral measurements. In this way the correct operation of this processed version is verified. In the second case unimodal distributions are obtained for all cases, probably due to the elimination of low level echoes.

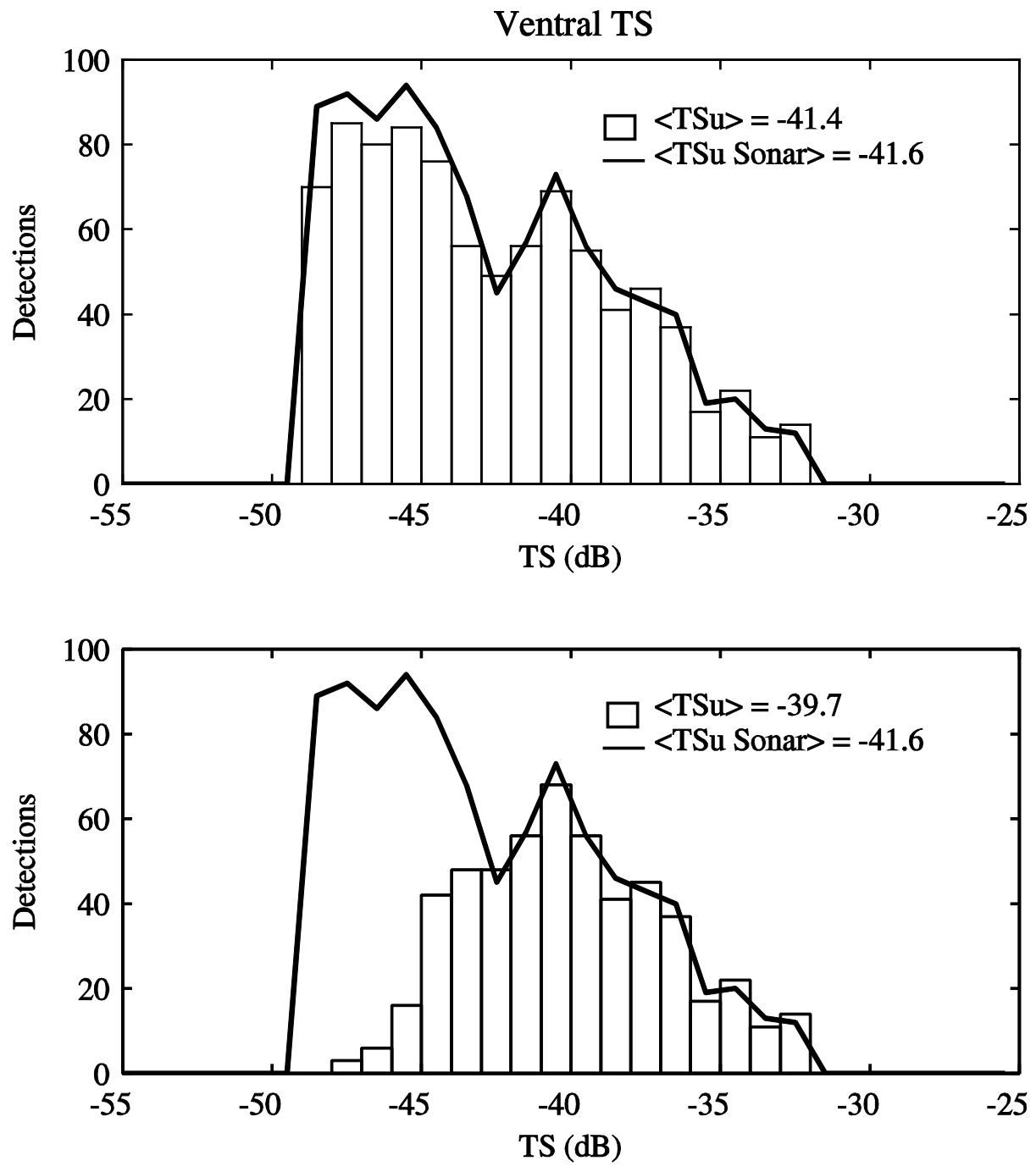

***Fig. 4***. *Ventral TSu distribution for size class 4 (23.1 cm) when echo length criterion (top) or threshold criterion (bottom) are applied firstly.*

A linear relationship between ventral *TSu* and logarithm of total body length of gilthead sea bream is observed. A good correlation between ventral aspect of *TSu* and total body length of fish is achieved, when the threshold criterion is applied first. In case of applying the echo length criterion firstly, a rather poor correlation and lower mean values of *TSu* are achieved, as lower level echoes are included (*Fig. 5*). Results do not provide a linear relationship for dorsal case, similar to the results observed for split-beam analysis.

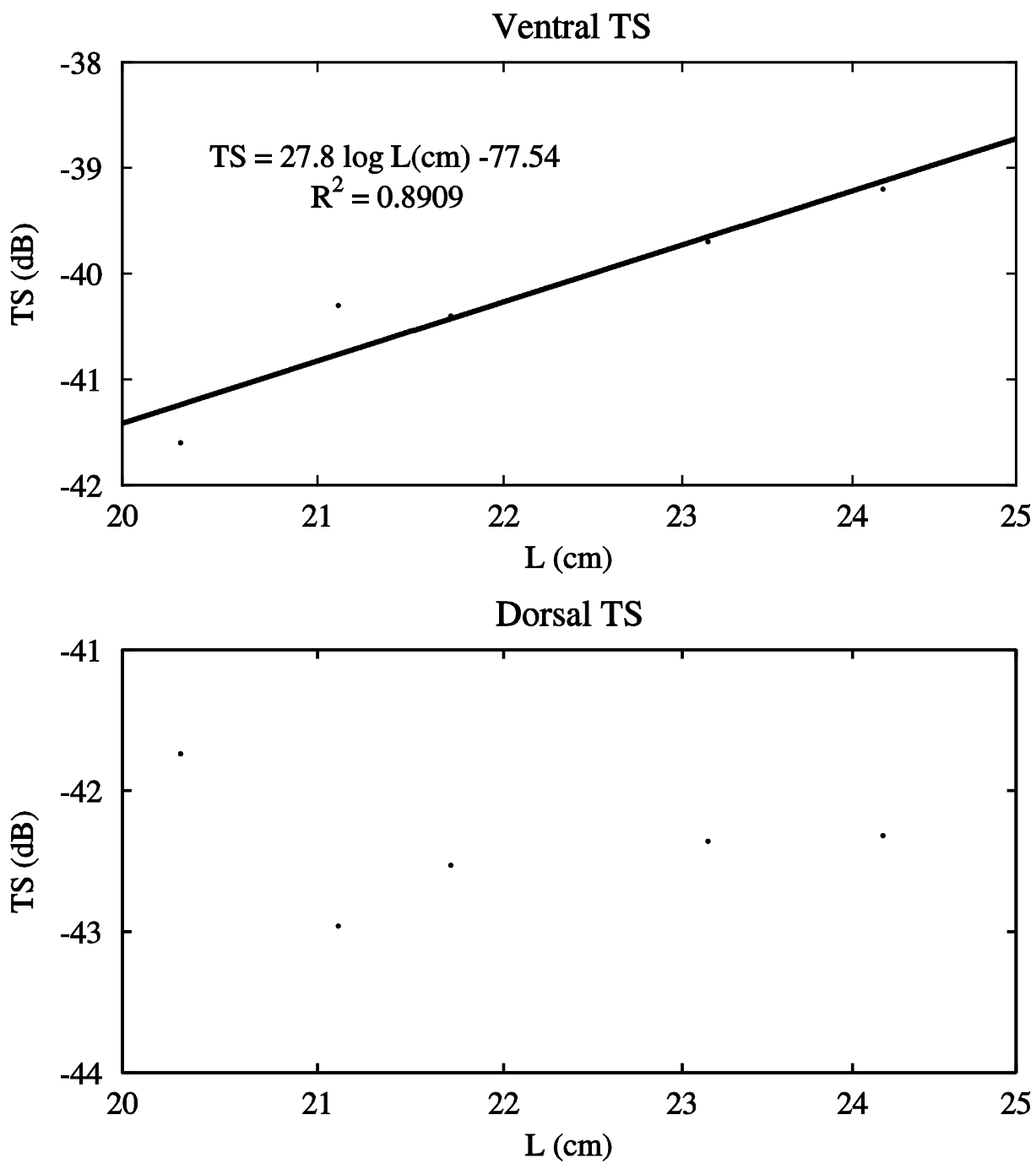

***Fig. 5.*** *Relationship between TSu and total body length (in centimeters) of the gilthead sea bream for both ventral (left) and dorsal aspect (right). TSu is obtained by applying the threshold criterion in first place.*

Presence of low level echoes or noise in size class 5 has not had the same influence as in the split-beam analysis. Application of the threshold criterion in the first place causes the elimination of such detections.

The following linear relationship between *TSu* ventral aspect and logarithm of total body length has been obtained at a frequency of 200 kHz,

$$TSu = 27.8 \cdot \log L\ (\mathrm{cm}) - 77.54 \qquad (2)$$

$$r^2 = 0.8909$$

# 4 Conclusions

No clear dependence of *TS* with range at short distances is observed. An empirical relationship has been obtained to estimate size of gilthead sea bream in sea cages from backscattering measurements, without making any correction for the effects at short distances (near field effect and violation of point source), as happens with other species (Knudsen et al., 2004).

However some single fish tracks have been found at short distances, even when fish length is greater than diameter of the acoustic beam. This may be because the swimbladder reflects 90% or more of backscattered energy (Simmonds and MacLennan, 2005), so it might be more reasonable to consider length of the swimbladder instead of total body length.

The swimbladder as dispersing element still occupies a large part of the acoustic beam cross section at short distances, so it is necessary to correct TVG function to obtain absolute *TS* values. Even though phase errors having an unknown magnitude, will occur.

*TS* distributions are unimodal for both ventral and dorsal measurements. This could be explained by a reduced directivity of specimens studied, because size of the swimbladder is only slightly higher than wavelength. Unimodal distributions are also obtained for *TSu*, when using the algorithm that applies first threshold criterion, therefore eliminating low-level echoes or noise.

*TS* for dorsal recordings is lower than ventral aspect of *TS*, like other species of fish (Knudsen et al., 2004), possibly due to shadowing effect of swimbladder by hard structures of fish. We have not found a clear relationship between *TS* dorsal aspect and logarithm of total body length of the gilthead sea bream.

The linear relationship obtained for ventral *TS* versus logarithm of total body length of the gilthead sea bream shows a good correlation. We found no clear relationship for dorsal measurements. Ventral *TS* for the largest class do not follow the trend of other classes, possibly due to the need of a greater number of measurements. Linear fit is not included due to this reason.

A linear relationship between *TS* and logarithm of total body length for ventral measurements has been obtained, as suggested by Ona (1999). A linear relationship is also found for *TSu*. In both cases the correlations obtained are quite good, despite the unavailability of a large number of measurements. These relationships enable to estimate fish size from a few simple steps and not extended in time. Note the importance of the relationship found between *TSu* and length of fish, allowing us to monitor the growth of gilthead sea bream in sea cages from a few simple measurements performed with a low-cost equipment. However, results are sensitive to processing parameters, therefore it is desirable to extend the analysis in future studies to a higher number of detections and a greater number of size classes, it may prove more conclusively intended use of *TSu* as a parameter for growth monitoring of gilthead sea bream in sea cages.

It should be taking into account that *TSu* is practically the *TS*. Since a high threshold has been applied in single-beam analysis, the echoes from off-axis have been removed. *TS* values are rather higher than *TSu* values, and higher slope. The lower values of *TSu* are expected, since transducer directivity pattern is not compensated. It has to be noticed strong difference in the slope between approaches, split and single-beam. For split-beam data the *TS* increases faster with fish length. An explanation can be given in terms of the possible over-compensation of beam directivity in split-beam analysis. This compensation is given for a determinate angular position, and since fishes occupy an extended area inside the beam, the bigger the fish the bigger the error in directivity compensation.

# Acknowledgements

Especially thank Javier Zaragozá for their help in the design and mounting of the sea cage, always supporting beyond duty, Andres Moñino (ICTA, UPV) by fish manipulation, and Helge Balk (Sonar5-Pro developer) for their valuable comments and constant support. We must also acknowledge the Autoridad Portuaria de Gandia (Valencia) collaboration and facilities for placing the sea cage.

# References

APROMAR, 2009. La acuicultura marina de peces en España 2010. http://www.apromar.es/Informes/Informe-APROMAR-2010.pdf

Balk, H., Lindem, T., 2009. Sonar4 and Sonar5-Pro post processing systems. Operator manual version 5.9.8. University of Oslo, Norway.

Dawson, J. J., Wiggins, D., Degan, D., Geiger, H., Hart, D., Adams, B., 2000. Point-source violations: split-beam tracking of fish at close range. Aquat. Living Resour. 13(5), 291-295.

Edwards, J. I., Amstrong, F., 1983. Measurement of the target strength of live herring and mackerel. FAO Fish. Rep. 300, 69-77.

FAO, 2010. Food and Agriculture Organization of the United Nations (FAO) - Fisheries and Aquaculture Department. http://www.fao.org/fishery/aquaculture/es

Foote, K. G., 1985. Rather-high-frequency sound scattered by swimbladdered fish. J. Acoust. Soc. Am. 78, 688-700.

Gauthier, S., Rose, G. A., 2001. Target strength of encaged Atlantic redfish (*Sebastes* spp.). ICES J. Mar. Sci. 58, 562-568.

Gauthier, S., Rose, G. A., 2002. An hypothesis on endogenous hydrostasis in Atlantic redfish (*Sebastes* spp.). Fish. Res. 58, 227-230.

Jorgensen, R., 2003. The effects of swimbladder size, condition and gonads on the acoustic target strength of mature capelin. ICES J. Mar. Sci. 60, 1056-1062.

Jorgensen, R., Olsen, K., 2002. Acoustic target strength of capelin measured by single-target tracking in a controlled cage experiment. ICES J. Mar. Sci. 59, 1081-1085.

Kang, D., Cho, S., Lee, C., Myoung, J.-G., Na, J., 2009. Ex situ target-strength measurements of Japanese anchovy (*Engraulis japonicus*) in the coastal Northwest Pacific. ICES J. Mar. Sci. 66, 1219-1224.

Kang, D., Hwang, D., 2003. Ex situ target strength of rockfish (*Sebastes schlegeli*) and red sea bream (*Pagrus major*) in the Northwest Pacific. ICES J. Mar. Sci. 60, 538-543.

Kang, D., Sadayasu, K., Mukai, T., Iida, K., Hwang, D., Sawada, K., Miyashita, K., 2004. Target strength estimation of black porgy Acanthopagrus schlegeli using acoustic measurements and a scattering model. Fish. Sci. 70, 819-828.

Knudsen, F. R., Fosseidengen, J. E., Oppedal, F., Karlsen, O., Ona, E., 2004. Hydroacoustic monitoring of fish in sea cages: target strength (*TS*) measurements on Atlantic salmon (*Salmo salar*). Fish. Res. 69, 205-209.

Lilja, J., Marjomaki, T. J., Riikonen, R., Jurvelius, J., 2000. Side-aspect targest strength of atlantic salmon (*Salmo salar*), brown trout (*Salmo trutta*), whitefish (*Coregonus lavaretus*), and pike (*Esox lucius*). Aquat. Living Resour. 13, 355-360.

McClatchie, S., Alsop, J., Coombs, R. F., 1996. A re-evaluation of relationships between fish size, acoustic frequency, and target strength. ICES J. Mar. Sci. 53, 780-791.

Medwin, H., 2005. Sounds in the sea: from ocean acoustic to acoustical oceanography, ed. Cambridge University Press, Cambridge

Medwin, H., Clay, C. S., 1998. Fundamentals of acoustical oceanography, ed. Academic Press, Boston.

Mukai, T., Iida, K., 1996. Depth dependence of target strength of live kokanee salmon in accordance with Boyle's law. ICES J. Mar. Sci. 53, 245-248.

Mulligan, T., 2000. Shallow water fisheries sonar: a personal view. Aquat. Living Resour. 13(5), 269-273.

Nielsen, J. R., Lundgren, B., 1999. Hydroacoustic ex situ target strength measurements on juvenile cod (*Gadus morhua* L.). ICES J. Mar. Sci. 56, 627-639.

Ona, E., 1999. Methodology for target strength measurements. ICES Corporative Res. Rep.nº 235.

Ona, E., 2003. An expanded target-strength relationship for herring. ICES J. Mar. Sci. 60, 493-499.

Simmonds, E. J., MacLennan, D. N., 2005. Fisheries acoustics: theory and practice, second ed. Blackwell Science, Oxford.